\documentclass[12pt]{article}
\usepackage{amssym}

\renewcommand{\theequation}{\arabic{section}.\arabic{equation}}
\newcommand{\xb}{\mbox{\boldmath $x$}}
\newcommand{\pb}{\mbox{\boldmath $p$}}
\newcommand{\pib}{\mbox{\boldmath $\pi$}}
\newcommand{\nablab}{\mbox{\boldmath $\nabla$}}
\newcommand{\xxb}{\mbox{\boldmath $X$}}
\newcommand{\ppb}{\mbox{\boldmath $P$}}
\newcommand{\pipib}{\mbox{\boldmath $\Pi$}}
\newcommand{\fb}{\mbox{\boldmath $f$}}
\newcommand{\ddb}{\mbox{\boldmath $D$}}
\newcommand{\llb}{\mbox{\boldmath $L$}}

\newcommand{\tpsi}{\tilde{\psi}}
\newcommand{\aalpha}{\overline{\alpha}}
\def\diver{\mathop{\rm div}\nolimits}
\def\N{\mbox{$\Bbb N$}}

\title{
Deformed algebras, position-dependent effective masses and curved spaces: An exactly
solvable Coulomb problem}
\author{C Quesne$^{\dagger}$ and  V M Tkachuk$^{\ddagger}$\\
$^{\dagger}$ {\small Physique Nucl\'eaire Th\'eorique et Physique
Math\'ematique,  Universit\'e Libre de Bruxelles,} \\ 
{\small Campus de la Plaine CP229, Boulevard~du Triomphe, B-1050 Brussels,
Belgium}\\ 
$^{\ddagger}$ {\small Ivan Franko Lviv National University, Chair of Theoretical
Physics,}\\
{\small 12, Drahomanov Street, Lviv UA-79005, Ukraine}\\
{\small E-mail: cquesne@ulb.ac.be and tkachuk@ktf.franko.lviv.ua}}
\date{ }
\begin{document}
\baselineskip=20pt plus 1pt minus 1pt
\maketitle
\begin{abstract}
We show that there exist some intimate connections between three unconventional
Schr\"odinger equations based on the use of deformed canonical commutation relations, of a
position-dependent effective mass or of a curved space, respectively. This occurs whenever a
specific relation between the deforming function, the position-dependent mass and the
(diagonal) metric tensor holds true. We illustrate these three equivalent approaches by
considering a new Coulomb problem and solving it by means of supersymmetric quantum
mechanical and shape invariance techniques. We show that in contrast with the conventional
Coulomb problem, the new one gives rise to only a finite number of bound states.
\end{abstract}

\vspace{0.5cm}


\newpage
\section{Introduction}

In recent years there has been a growing interest in the study of quantum mechanical systems
with a position-dependent effective mass due to their applications in condensed-matter
physics. Initially proposed to describe impurities in crystals~\cite{wannier}, the effective-mass
theory has become an essential ingredient in the description of electronic properties of
semiconductors~\cite{bastard} and quantum dots~\cite{serra}, for instance.\par
%
%
The concept of effective mass is also relevant in connection with the energy-density
functional approach to the quantum many-body problem. This formalism has been extensively
used in nuclei~\cite{ring}, quantum liquids~\cite{arias}, ${}^3$He clusters~\cite{barranco},
and metal clusters~\cite{puente}.\par
%
%
The study of quantum mechanical systems with position-dependent mass raises some
important conceptual problems, such as the ordering ambiguity of the momentum and mass
operators in the kinetic energy term, the boundary conditions at abrupt interfaces
characterized by discontinuities in the mass function, and the Galilean invariance of the theory
(see, e.g.,~\cite{levy, cavalcante}).\par
%
%
In the standard case of constant mass, exactly solvable (ES) Schr\"odinger equations have
played an important role because they provide both a conceptual understanding of some
physical phenomena and a testing ground for some approximation schemes. Many different
approaches have been used including point canonical transformations (PCT)~\cite{bhatta},
Lie algebraic methods~\cite{alhassid}, supersymmetric quantum mechanical (SUSYQM) and
shape invariance (SI) techniques~\cite{gendenshtein}. The class of ES potentials has also
been extended to the so-called quasi-exactly solvable (QES) potentials~\cite{turbiner} and
the conditionally exactly solvable (CES) ones~\cite{dutra93}. During the last few years,
some of these developments have been generalized to the case of position-dependent
mass~\cite{dekar, milanovic, plastino, dutra00, roy, koc, alhaidari, gonul}.\par
%
%
The main purpose of the present paper is to establish some connections between the
Schr\"odinger equation with position-dependent mass and two other extensions of the
conventional Schr\"odinger equation in current use.\par
%
%
The first one is based on the replacement of the standard commutation relations by deformed
ones. A motivation for such an approach is the possibility of describing, for a special choice of
deformation, nonzero minimal uncertainties in position and/or momentum~\cite{kempf94}.
This is in line with several investigations in string theory and quantum gravity, which suggest the
existence of a finite lower bound to the possible resolution of length and/or momentum (see,
e.g.,~\cite{witten}). In this context, several ES problems related to the harmonic oscillator
have recently been considered using either the PCT~\cite{kempf95} or the SUSYQM~\cite{cq}
method.\par
%
%
The second extension has to do with the Schr\"odinger equation in curved space. The interest
in such a problem, which dates back to Schr\"odinger himself~\cite{schrodinger}, has found a
revival with the advent of deformed algebras~\cite{higgs, granovskii}. Since then, many
studies have been devoted to this topic, especially in the case of spaces of constant
curvature (see, e.g.,~\cite{kalnins96}), but also in the more general case of nonconstant
curvature (see, e.g.,~\cite{kalnins02}).\par
%
%
Another purpose of the present paper is to illustrate these three equivalent approaches by
considering a new Coulomb problem with a definite dependence of the mass on the radial
variable. We plan to show that the corresponding Schr\"odinger equation is ES by using the
same kind of SUSYQM approach as that recently applied to the harmonic oscillator problem
with nonzero minimal uncertainties in position and/or momentum~\cite{cq}.\par
%
%
Our paper is organized as follows. Connections between some unconventional Schr\"odinger
equations are established in section~2. Section~3 deals with a new Coulomb problem. Finally,
section~4 contains the conclusion.\par
%
%
\section{Connections between some unconventional Schr\"odinger equations}

In the present section, we will successively consider the cases where the Schr\"odinger
equation is used in combination with some deformed canonical commutation relations or it
contains a position-dependent mass or else the underlying space is curved and we will
establish some connections between these three approaches. For definiteness sake, we will
restrict ourselves here to the example of a single-particle in three dimensions, but it should be
clear that similar relations exist in more general cases.\par
%
%
\subsection{Deformed canonical commutation relations}

In the conventional canonical commutation relations
\begin{eqnarray}
  [x_i, x_j] & = & 0  \label{eq:com-xx}\\{}
  [x_i, p_j] & = & {\rm i} \hbar \delta_{i,j}  \label{eq:com-xp}\\{}
  [p_i, p_j] & = & 0  \label{eq:com-pp}
\end{eqnarray}
where $i=1,$ 2, 3, let us replace the momentum components $p_i = - {\rm i} \hbar \nabla_i
= - {\rm i} \hbar \partial/\partial x_i$ by some deformed Hermitian operators
\begin{equation}
  \pi_i = \sqrt{f(\xb)}\, p_i \sqrt{f(\xb)}  \label{eq:pi}
\end{equation}
where the (positive real) deforming function $f$ depends on the coordinates $\xb = (x_1, x_2, x_3)$. It
is then straightforward to show that while equation (\ref{eq:com-xx}) still holds, equations
(\ref{eq:com-xp}) and (\ref{eq:com-pp}) are replaced by 
\begin{eqnarray}
  [x_i, \pi_j] & = & {\rm i} \hbar f(\xb) \delta_{i,j} \label{eq:com-xpi} \\{}
  [\pi_i, \pi_j] & = & - {\rm i} \hbar [f_i(\xb) \pi_j - f_j(\xb) \pi_i] \label{eq:com-pipi}
\end{eqnarray}
where $f_i(\xb) \equiv \nabla_i f(\xb)$.\par
%
%
In the special case where $f$ only depends on the radial variable $r = \left(\sum_i
x_i^2\right)^{1/2}$, equations (\ref{eq:com-xpi}) and (\ref{eq:com-pipi}) become
\begin{eqnarray}
  [x_i, \pi_j] & = & {\rm i} \hbar f(r) \delta_{i,j} \\{}
  [\pi_i, \pi_j] & = & - {\rm i} \hbar \frac{f(r)f'(r)}{r}\, \epsilon_{ijk} l_k  \label{eq:com-pipi-r}
\end{eqnarray}
where $f'(r) = df(r)/dr$, $\epsilon_{ijk}$ is the antisymmetric tensor and
\begin{equation}
  l_i = \epsilon_{ijk} x_j p_k
\end{equation}
are the angular momentum components. The latter satisfy the usual commutation relations
with $x_i$ and $\pi_i$, i.e.,
\begin{equation}
  [l_i, x_j] = {\rm i} \hbar \epsilon_{ijk} x_k \qquad [l_i, \pi_j] = {\rm i} \hbar \epsilon_{ijk}
  \pi_k
\end{equation}
from which it follows that the relation $[l_i, f(r)] = 0$ holds true. Such a property has actually
been used in the derivation of (\ref{eq:com-pipi-r}).\par
%
%
Let us now consider a deformed Schr\"odinger operator
\begin{equation}
  H_1 = \frac{1}{2m_0} \pib^2 + V_1(\xb)  \label{eq:H-1} 
\end{equation}
where $m_0$ denotes the (constant) mass and $V_1(\xb)$ is some potential. The
corresponding deformed Schr\"odinger equation reads
\begin{equation}
  \left[- \frac{\hbar^2}{2m_0} \left(\sqrt{f(\xb)}\, \nablab \sqrt{f(\xb)}\right)^2 +
  V_1(\xb)\right] \psi(\xb) = E \psi(\xb).  \label{eq:SE-1} 
\end{equation}
\par
%
%
The problem can be reformulated in terms of dimensionless operators $\xxb = \xb/a$, $\ppb
= a \pb/\hbar$, $\pipib = a \pib/\hbar$, where $a$ denotes some characteristic length.
Such new operators satisfy commutation relations similar to (\ref{eq:com-xx}),
(\ref{eq:com-xp}), (\ref{eq:com-pp}) [or (\ref{eq:com-xx}), (\ref{eq:com-xpi}),
(\ref{eq:com-pipi})] where $\hbar$ is set equal to 1 and $l_i$ is replaced by $L_i =
\epsilon_{ijk} X_j P_k$. A dimensionless Hamiltonian can be defined by
\begin{equation}
  h_1 = \frac{H_1}{V_0} = \frac{1}{2} \pipib^2 + U_1(\xxb)
\end{equation}
where $V_1(\xb) = V_0 U_1(\xxb)$ and $V_0 = \hbar^2/(m_0 a^2)$. Equation
(\ref{eq:SE-1}) then becomes
\begin{equation}
  \left[- \frac{1}{2} \left(\sqrt{f(\xxb)}\, \nablab \sqrt{f(\xxb)}\right)^2 +
  U_1(\xxb)\right] \psi(\xxb) = e \psi(\xxb)  \label{eq:SE-1bis} 
\end{equation}
where $e = E/V_0$ and for simplicity's sake, we keep the notation $\nabla_i$ to denote the
derivatives $\partial/\partial X_i$ with respect to the dimensionless variables $X_i$.\par
%
%
\subsection{Position-dependent mass}

When the mass $m(\xb)$ is position dependent, it does no longer commute with the
momentum $\pb = - {\rm i} \hbar \nablab$, so that there are many ways of generalizing the
usual form of the kinetic energy $(2m_0)^{-1} \pb^2$, valid for a constant mass $m_0$, in
order to obtain a Hermitian operator. This ordering ambiguity has been most debated (see,
e.g.,~\cite{levy, cavalcante}). Here we will not make any specific choice and will therefore
adopt the general two-parameter form of the kinetic energy term, as originally proposed by
von Roos~\cite{vonroos}.\par
%
%
Hence we consider a Hamiltonian $H_2$ defined by
\begin{equation}
  H_2 = - \frac{\hbar^2}{4} \left[m^{\delta'}(\xb) \nablab m^{\kappa'}(\xb) \nablab
  m^{\lambda'}(\xb) + m^{\lambda'}(\xb) \nablab m^{\kappa'}(\xb) \nablab
  m^{\delta'}(\xb)\right] + V_2(\xb)  \label{eq:H-2} 
\end{equation}
where $V_2(\xb)$ is some potential and the parameters $\delta'$, $\kappa'$, $\lambda'$
are constrained by the condition $\delta' + \kappa' + \lambda' = -1$. On expressing the
position-dependent mass $m(\xb)$ as
\begin{equation}
  m(\xb) = m_0 M(\xb) \qquad M(\xb) = \frac{1}{f^2(\xb)}  \label{eq:mass}
\end{equation}
where $m_0$ is a constant mass and $M(\xb)$ is a dimensionless position-dependent mass, 
equation (\ref{eq:H-2}) can be rewritten as
\begin{equation}
  H_2 = - \frac{\hbar^2}{4m_0} \left[f^{\delta}(\xb) \nablab f^{\kappa}(\xb) \nablab
  f^{\lambda}(\xb) + f^{\lambda}(\xb) \nablab f^{\kappa}(\xb) \nablab
  f^{\delta}(\xb)\right] + V_2(\xb)  \label{eq:H-2-bis} 
\end{equation}
with $\delta + \kappa + \lambda = 2$. For the special choice $\delta' = \lambda' = 0$ and
$\kappa' = -1$ or $\delta = \lambda = 0$ and $\kappa = 2$, equations (\ref{eq:H-2}) and
(\ref{eq:H-2-bis}) reduce to the most common BenDaniel-Duke form~\cite{bendaniel}
\begin{equation}
  H'_2 = - \frac{\hbar^2}{2} \nablab \frac{1}{m(\xb)} \nablab + V_2(\xb) = - \frac{\hbar^2}{2m_0}
  \nablab f^2(\xb) \nablab + V_2(\xb).
\end{equation}
\par
%
%
We now plan to show that the Hamiltonian (\ref{eq:H-2-bis}) can be transformed into
\begin{equation}
  H_2 = - \frac{\hbar^2}{2m_0} \sqrt{f(\xb)}\, \nablab f(\xb) \nablab \sqrt{f(\xb)} + 
  V_{2,{\rm eff}}(\xb)  \label{eq:H-2-ter}
\end{equation}
for some appropriate choice of the effective potential $V_{2,{\rm eff}}(\xb)$. Since
\begin{eqnarray}
  && f^{\delta} \nablab f^{\kappa} \nablab f^{\lambda} \nonumber\\
  && = f^{\delta} \nablab f^{\frac{1}{2} - \delta} f f^{\frac{1}{2} - \lambda} \nablab
       f^{\lambda}  \nonumber \\
  && = \left[f^{1/2} \nablab + \left(\frac{1}{2} - \delta\right) f^{-1/2} \fb\right] f  
       \left[\nablab f^{1/2} - \left(\frac{1}{2} - \lambda\right) f^{-1/2} \fb\right]
       \nonumber \\
  && = \sqrt{f}\, \nablab f \nablab \sqrt{f} + (\lambda - \delta) \sqrt{f}\,  \fb \cdot \nablab
       \sqrt{f} - \left(\frac{1}{2} - \lambda\right) f \diver \fb \nonumber \\
  && \hphantom{=} - \left(\frac{1}{2} - \delta\right) \left(\frac{1}{2} - \lambda\right)
       \fb^2  \label{eq:lemma}  
\end{eqnarray}
with $\fb = \nablab f$, we indeed obtain
\begin{eqnarray}
  \lefteqn{f^{\delta} \nablab f^{\kappa} \nablab f^{\lambda} + f^{\lambda} \nablab
       f^{\kappa} \nablab f^{\delta}}  \nonumber \\
  && = 2 \sqrt{f}\, \nablab f \nablab \sqrt{f}  - (1 - \delta - \lambda) f \diver \fb
       - 2 \left(\frac{1}{2} - \delta\right) \left(\frac{1}{2} - \lambda\right) \fb^2.  
\end{eqnarray}
We conclude that equation (\ref{eq:H-2-ter}) is valid for
\begin{equation}
  V_{2,{\rm eff}}(\xb) = V_2(\xb) + \frac{\hbar^2}{2m_0} \left[\frac{1}{2} (1 - \delta -
  \lambda) f(\xb) \diver \fb(\xb) + \left(\frac{1}{2} - \delta\right) \left(\frac{1}{2} -
\lambda\right) \fb^2(\xb)\right].  \label{eq:V-2eff}
\end{equation}
\par
%
%
On taking equation (\ref{eq:pi}) into account, it becomes clear that the Hamiltonian $H_2$
with position -dependent mass (\ref{eq:mass}) may be considered as a deformed
Schr\"odinger Hamiltonian $H_1$, as given in (\ref{eq:H-1}), with $V_1(\xb) = V_{2,{\rm
eff}}(\xb)$.\par
%
%
In terms of dimensionless variables $\xxb = \xb/a$, there corresponds to $H_2$ a
dimensionless Hamiltonian
\begin{equation}
  h_2 = \frac{H_2}{V_0} = - \frac{1}{2} \sqrt{f(\xxb)}\, \nablab f(\xxb) \nablab
  \sqrt{f(\xxb)} + U_{2,{\rm eff}}(\xxb)
\end{equation}
and a corresponding Schr\"odinger equation
\begin{equation}
  \left[- \frac{1}{2} \left(\sqrt{f(\xxb)}\, \nablab \sqrt{f(\xxb)}\right)^2 +
  U_{2,{\rm eff}}(\xxb)\right] \psi(\xxb) = e \psi(\xxb)  \label{eq:SE-2} 
\end{equation}
where $U_{2,{\rm eff}}(\xxb) = V_{2,{\rm eff}}(\xb)/V_0$, $e = E/V_0$, and $V_0 =
\hbar^2/(m_0 a^2)$. We conclude that solving the deformed Schr\"odinger equation
(\ref{eq:SE-1bis}) will also provide us with the solutions of equation (\ref{eq:SE-2}) provided
\begin{equation}
  U_{2,{\rm eff}}(\xxb) = U_1(\xxb).
\end{equation}
\par
%
%
In the special case of a central potential $V(r)$ with a mass depending only on the radial
variable $r$, i.e., $m(r) = m_0 M(r) = m_0/f^2(r)$, the effective potential (\ref{eq:V-2eff})
reduces to
\begin{equation}
  V_{2,{\rm eff}}(r) = V_2(r) + \frac{\hbar^2}{2m_0} \left\{\frac{1}{2} (1 - \delta -
  \lambda) f(r) \left[\frac{2}{r} f'(r) + f''(r)\right] + \left(\frac{1}{2} - \delta\right)
  \left(\frac{1}{2} - \lambda\right) f^{\prime2}(r)\right\}.  
\end{equation}
In terms of dimensionless quantities, this becomes
\begin{equation}
  U_{2,{\rm eff}}(\rho) = U_2(\rho) + \frac{1}{2} \left\{\frac{1}{2} (1 - \delta - \lambda)
  f(\rho) \left[\frac{2}{\rho} f'(\rho) + f''(\rho)\right] + \left(\frac{1}{2} - \delta\right)
  \left(\frac{1}{2} - \lambda\right) f^{\prime2}(\rho)\right\} 
\end{equation}
for a mass $m(\rho) = m_0 M(\rho) = m_0/f^2(\rho)$, where $\rho = r/a$.\par
%
%
\subsection{Curved space}

Let us consider a curved space, whose squared line element is given by
\begin{equation}
  ds^2 = g_{ij}(\xb) dx^i dx^j  \label{eq:line}
\end{equation}
where the metric tensor is assumed to be of the form
\begin{equation}
  g_{ij}(\xb) = g_{ii}(\xb) \delta_{i,j} = D_i^2(\xb) \delta_{i,j}
\end{equation}
with $g_{ii}(\xb) = D_i^2(\xb)$ independent of $i$, i.e.,
\begin{equation}
  g_{ii}(\xb) = g(\xb) \qquad D_i(\xb) = D(\xb).
\end{equation}
\par
%
%
The corresponding Laplacian operator~\cite{gradshteyn}
\begin{equation}
  \Delta = \frac{1}{D_1D_2D_3} \left(\frac{\partial}{\partial x_1} \frac{D_2D_3}{D_1}
  \frac{\partial}{\partial x_1} + \frac{\partial}{\partial x_2} \frac{D_1D_3}{D_2}
  \frac{\partial}{\partial x_2} + \frac{\partial}{\partial x_3} \frac{D_1D_2}{D_3}
  \frac{\partial}{\partial x_3}\right)
\end{equation}
then reduces to
\begin{equation}
  \Delta = \frac{1}{D^3(\xb)} \nablab D(\xb) \nablab = \nablab \frac{1}{D^2(\xb)}
  \nablab + \frac{3}{D^3(\xb)} \ddb(\xb) \cdot \nablab
\end{equation}
where $\nabla_i = \partial/\partial x_i$ and $\ddb(\xb) = \nablab D(\xb)$. On expressing
$g(\xb)$ or $D(\xb)$ as 
\begin{equation}
  g(\xb) = \frac{1}{f^2(\xb)} \qquad D(\xb) = \frac{1}{f(\xb)}
\end{equation}
$\Delta$ can be rewritten as
\begin{equation}
  \Delta = \nablab f^2(\xb) \nablab - 3 f(\xb) \fb(\xb) \cdot \nablab.  \label{eq:laplacian}
\end{equation}
\par
%
%
In such a curved space, let us now consider some Hamiltonian
\begin{equation}
  H_3 = - \frac{\hbar^2}{2m_0} \Delta + V_3(\xb)  \label{eq:H-3}
\end{equation}
where $m_0$ denotes a (constant) mass and $V_3(\xb)$ is some potential. There
corresponds to it a dimensionless Hamiltonian $h_3 = H_3/V_0$, depending on a
dimensionless potential $U_3(\xxb) = V_3(\xb)/V_0$, where $\xxb = \xb/a$ and $V_0 =
\hbar^2/(m_0 a^2)$.\par
%
%
{}From equation (\ref{eq:laplacian}), it follows that the corresponding Schr\"odinger equation
reads
\begin{equation}
  \left[- \frac{1}{2} \nablab f^2(\xxb) \nablab + \frac{3}{2} f(\xxb) \fb(\xxb) \cdot
  \nablab + U_3(\xxb)\right] \tpsi(\xxb) = e \tpsi(\xxb)  \label{eq:SE-3}
\end{equation}
where here $\nabla_i = \partial/\partial X_i$. The second term on the left-hand side
containing first-order derivatives can be eliminated by setting
\begin{equation}
  \tpsi(\xxb) = [f(\xxb)]^{3/2} \psi(\xxb).  \label{eq:change}
\end{equation}
The resulting equation can be written as
\begin{equation}
  \left\{- \frac{1}{2} \nablab f^2(\xxb) \nablab + \frac{3}{4}\left[\frac{1}{2}\fb^2(\xxb) -
  f(\xxb) \diver \fb(\xxb)\right] + U_3(\xxb)\right\} \psi(\xxb) = e \psi(\xxb)
\end{equation}
or, alternatively,
\begin{equation}
  \left[- \frac{1}{2}\left(\sqrt{f(\xxb}\, \nablab \sqrt{f(\xxb)}\right)^2 +
  U_{3,{\rm eff}}(\xxb)\right] \psi(\xxb) = e \psi(\xxb)  \label{eq:SE-3bis}
\end{equation}
where
\begin{equation}
  U_{3,{\rm eff}}(\xxb) = U_3(\xxb) - \frac{1}{2} f(\xxb) \diver \fb(\xxb) + \frac{1}{2}
  \fb^2(\xxb).  \label{eq:U-3eff}
\end{equation}
Note that in the last step, we used equation (\ref{eq:lemma}) with $\delta = \lambda = 0$
and $\kappa = 2$.\par
%
%
We have therefore proved that solving the deformed Schr\"odinger equation
(\ref{eq:SE-1bis}) also provides us with the solutions of equation (\ref{eq:SE-3}) provided we
make the identification
\begin{equation}
  U_{3,{\rm eff}}(\xxb) = U_1(\xxb)
\end{equation}
and take equations (\ref{eq:change}) and (\ref{eq:U-3eff}) into account.\par
%
%
Note that whenever the potential $U_3$ and the metric tensor $g = 1/f^2$ only depend on
$\rho = \left(\sum_i X_i^2\right)^{1/2}$, the effective potential reduces to
\begin{equation}
  U_{3,{\rm eff}}(\rho) = U(\rho) - \frac{1}{2} f(\rho) \left(\frac{2}{\rho} f'(\rho) + f''(\rho)
  \right) + \frac{1}{2} f^{\prime2}(\rho).
\end{equation}
\par
%
%
As shown in the appendix, in such a special case, the space curvature takes a simple form in
terms of the metric $g(\rho)$ or the deforming function $f(\rho)$,
\begin{equation}
  R = 2 \left(- \frac{4}{\rho} f(\rho) f'(\rho) - 2 f(\rho) f''(\rho) + 3 f^{\prime2}(\rho)
  \right).  \label{eq:curvature}
\end{equation}
From this equation, it is clear that for a generic choice of $f(\rho)$, $R$ is not a constant, but
some function of $\rho$.\par
%
%
\bigskip\bigskip
We can summarize the results of this section as follows: we have established some closed
links between the deformed Schr\"odinger equation, the Schr\"odinger equation with
position-dependent mass and the Schr\"odinger equation in curved space whenever the
deforming function $f(\xb)$, the (dimensionless) position-dependent mass $M(\xb)$ and the
(diagonal) metric $g(\xb)$ are connected through the relations
\begin{equation}
  f^2(\xb) = \frac{1}{M(\xb)} = \frac{1}{g(\xb)}.  \label{eq:result}
\end{equation}
\par
%
%
\section{An exactly solvable Coulomb problem}
\setcounter{equation}{0}

Let us illustrate the general results obtained in section~2 by considering a deforming function
and a potential energy depending only on $r$ and given by
\begin{equation}
  f(r) = 1 + \aalpha r \qquad V(r) = - \frac{Z e^2}{r}  \label{eq:choice} 
\end{equation}
respectively. Here $\aalpha$ is some nonnegative parameter, $Z$ the atomic number and $e$
the electronic charge. For the Coulomb potential, the characteristic length is the Bohr radius
$a = \hbar^2/(m_0 e^2)$, where $m_0$ denotes the (undressed) mass of the electron. It
follows that $V_0 = \hbar^2/(m_0 a^2) = m_0 e^4/\hbar^2 = 2 R$, where $R$ is the
Rydberg constant.\par
%
%
In terms of the dimensionless radial variable $\rho = r/a$, equation (\ref{eq:choice}) can be
rewritten as 
\begin{equation}
  f(\rho) = 1 + \alpha \rho \qquad U(\rho) = - \frac{Z}{\rho}  \label{eq:choice-bis}
\end{equation}
where $\alpha = a \aalpha$.\par
%
%
\subsection{Deformed canonical commutation relations}

{}For the choice made in equation (\ref{eq:choice-bis}), the deformed canonical commutation
relations satisfied by the dimensionless operators $X_i$, $\Pi_i$ read
\begin{eqnarray}
  [X_i, X_j] & = & 0  \\{}
  [X_i, \Pi_j] & = & {\rm i} (1 + \alpha \rho) \delta_{i,j} \\{}
  [\Pi_i, \Pi_j] & = & -{\rm i} (1 + \alpha \rho) \frac{\alpha}{\rho} \epsilon_{ijk} L_k  
\end{eqnarray}
while, in spherical coordinates $\rho$, $\theta$, $\varphi$, the deformed Schr\"odinger
equation (\ref{eq:SE-1bis}) becomes
\begin{equation}
  \left\{- \frac{1}{2} \sqrt{f(\rho)} \left[f(\rho) \left(\frac{\partial^2}{\partial\rho^2} +
  \frac{2}{\rho} \frac{\partial}{\partial \rho} - \frac{\llb^2}{\rho^2}\right) + f'(\rho)
  \frac{\partial}{\partial \rho}\right] \sqrt{f(\rho)} - \frac{Z}{\rho}\right\} \psi(\rho, \theta,
  \varphi) = e \psi(\rho, \theta, \varphi)  \label{eq:SE-1C}
\end{equation}
where $\llb^2$ denotes the square of the angular momentum operator. In deriving equation
(\ref{eq:SE-1C}), we have used the relation
\begin{equation}
  \nablab f(\rho) \nablab = f(\rho) \nablab^2 + \frac{f'(\rho)}{\rho} \xxb \cdot \nablab. 
\end{equation}
\par
%
%
As in the usual $f(\rho)=1$ case, equation (\ref{eq:SE-1C}) is separable. On setting
\begin{equation}
  \psi_{klm}(\rho, \theta, \varphi) = \frac{1}{\rho} R_{kl}(\rho) Y_{lm}(\theta, \varphi)
  \label{eq:psi}
\end{equation}
where $Y_{lm}(\theta, \varphi)$ is a spherical harmonics, we get the radial differential
equation
\begin{equation}
  h^{(l)}_1 R_{kl}(\rho) = e_{kl} R_{kl}(\rho).  \label{eq:SE-1Cbis}
\end{equation}
Here
\begin{eqnarray}
  h^{(l)}_1 & = & - \frac{1}{2} \left(\sqrt{f(\rho)}\, \frac{d}{d\rho} \sqrt{f(\rho)}\right)^2 +
        U^{(l)}_1(\rho)  \\
  U^{(l)}_1(\rho) & = & \frac{1}{2} \left(f^2(\rho) \frac{L^2}{\rho^2} + \frac{f(\rho)
        f'(\rho)}{\rho}\right) - \frac{Z}{\rho} \nonumber \\
  & = & \frac{1}{2} \left(- \frac{2Z - \alpha(2L^2+1)}{\rho} + \frac{L^2}{\rho^2} + \alpha^2
        (L^2+1)\right)
\end{eqnarray}
where
\begin{equation}
  L^2 = l(l+1)  \label{eq:L-2}
\end{equation}
is the eigenvalue of $\llb^2$ and $k$ denotes the radial quantum number.\par
%
%
We now plan to show that equation (\ref{eq:SE-1Cbis}) is ES by applying the same kind of
SUSYQM methods as we used in~\cite{cq} to solve the harmonic oscillator problem with
nonzero minimal uncertainties in position and/or momentum.\par
%
%
To start with, the radial Hamiltonian $h^{(l)}_1$ can be factorized as
\begin{equation}
  h^{(l)}_1 = B^+(g, s) B^-(g, s) + \epsilon_0
\end{equation}
where the first-order operators $B^{\pm}(g, s)$ and the factorization energy $\epsilon_0$
are given by
\begin{eqnarray}
  B^{\pm}(g, s) & = & \frac{1}{\sqrt{2}}\left(\mp \sqrt{f(\rho)}\, \frac{d}{d\rho}
       \sqrt{f(\rho)} - \frac{s}{\rho} + g\right) \\
  s & = & l+1 \qquad g = \frac{Z - \frac{\alpha}{2}[(l+1)^2+L^2+1]}{l+1}  \label{eq:s-g} \\
  \epsilon_0 & = & - \frac{1}{2} g^2 + \frac{1}{2} \alpha^2 (L^2+1)  \label{eq:epsilon-0}
\end{eqnarray}
respectively. In the $\alpha \to 0$ limit, we get $s = l+1$, $g = Z/(l+1)$ and $\epsilon_0 = -
Z^2/[2(l+1)^2]$, which correspond to the usual factorization for the Coulomb potential in
conventional quantum mechanics~\cite{gendenshtein}.\par
%
%
Let us next consider a hierarchy of Hamiltonians
\begin{equation}
  h^{(l)}_{1i} = B^+(g_i, s_i) B^-(g_i, s_i) + \sum_{j=0}^i \epsilon_j \qquad i = 0, 1, 2, \ldots
\end{equation}
whose first member $h^{(l)}_{10}$ coincides with $h^{(l)}_1$ (hence $g_0 = g$ and $s_0
= s$), and let us impose a SI condition~\cite{gendenshtein}
\begin{equation}
  B^-(g_i, s_i) B^+(g_i, s_i) = B^+(g_{i+1}, s_{i+1}) B^-(g_{i+1}, s_{i+1}) + \epsilon_{i+1}.
  \label{eq:SI} 
\end{equation}
It can be easily shown that equation (\ref{eq:SI}) is satisfied provided
\begin{eqnarray}
  s_i & =& s + i = l + i + 1  \label{eq:s-i} \\
  g_i & = & \frac{g(l+1) - \frac{\alpha}{2}[l+1 + (2l+3)i + i^2]}{l+i+1} + \frac{\alpha}{2}
       \nonumber \label{eq:g-i} \\
  & =& \frac{Z - \frac{\alpha}{2}[(l+i+1)^2 + L^2 + 1]}{l+i+1} \\
  \epsilon_i & = & \frac{1}{2} g_{i-1}^2 - \frac{1}{2} g_i^2  \label{eq:epsilon-i}   
\end{eqnarray}
for $i=1$, 2,~\ldots. Again in the $\alpha \to 0$ limit, we get back the usual result $s_i =
l+i+1$ and $g_i = Z/(l+i+1)$.\par
%
%
The energy eigenvalues in equation (\ref{eq:SE-1Cbis}) can now be obtained from equations
(\ref{eq:epsilon-0}) and (\ref{eq:epsilon-i}) as
\begin{equation}
  e_{kl} = e_k(g, s) = \sum_{i=0}^k \epsilon_i = - \frac{1}{2} g_k^2 + \frac{1}{2} \alpha^2
  (L^2+1).  \label{eq:eigenvalue}
\end{equation}
Inserting equation (\ref{eq:g-i}) in equation (\ref{eq:eigenvalue}) converts the latter into
\begin{equation}
  e_{kl} = - \frac{1}{2} \left\{\frac{Z - \frac{\alpha}{2}[(l+k+1)^2 + L^2 + 1]}{l+k+1}
  \right\}^2 + \frac{1}{2} \alpha^2 (L^2+1)
\end{equation}
where $L^2$  is given in (\ref{eq:L-2}). In terms of the principal quantum number $n =
k+l+1$, the eigenvalues can be rewritten as
\begin{equation}
  e_{nl} = - \frac{\left[Z - \frac{\alpha}{2}(L^2+1)\right]^2}{2n^2} - \frac{\alpha^2}{8} n^2
  + \frac{\alpha}{2} \left[Z + \frac{\alpha}{2} (L^2+1)\right]
\end{equation}
which, in the $\alpha \to 0$ limit, leads to the usual result $e_{nl} = - Z^2/(2n^2)$. Note
that for $\alpha \ne 0$, there is an additional quadratic term in $n$, as well as an additional
dependence on $L^2$.\par
%
%
This purely algebraic determination of the spectrum has now to be completed by a
construction of the corresponding radial wave functions $R_{kl}(\rho)$, which should be
normalizable on $(0, \infty)$ according to~\footnote{It should be noted that in contrast
with~\cite{cq}, the scalar product is not changed by the deformation. With respect to
(\ref{eq:scal-prod}), the properties $[B^+(g, s)]^{\dagger} = B^-(g, s)$ and
$\bigl(h^{(l)}_1\bigr)^{\dagger} = h^{(l)}_1$ hold true.}
\begin{equation}
  \int_0^\infty d\rho\,  |R_{kl}(\rho)|^2 = 1.  \label{eq:scal-prod}
\end{equation}
As we now plan to show, this restricts the allowed (integer) values of $l$ and $k$ in contrast
with the standard Coulomb problem for which they may take any value in \N.\par
%
%
Let us first consider the ground state wave function $R_{0l}(\rho) = R_0(g, s; \rho)$ of
$h^{(l)}_1$, which is a solution of the first-order differential equation
\begin{equation}
  B^-(g, s) R_0(g, s; \rho) = 0.
\end{equation}
It is given by
\begin{equation}
  R_0(g, s; \rho) = {\cal N}_0(g, s) \rho^s (1 + \alpha\rho)^{-\left(\frac{g}{\alpha} + s +
  \frac{1}{2}\right)}
\end{equation}
which is a square-integrable function provided
\begin{equation}
  s > 0 \qquad g > 0.  \label{eq:condition1}
\end{equation}
In such a case, the normalization constant ${\cal N}_0(g, s)$ is
\begin{equation}
  {\cal N}_0(g, s) = \left(\frac{\Gamma\left(2\frac{g}{\alpha} + 2s + 1\right)}
  {\Gamma\left(2\frac{g}{\alpha}\right) \Gamma(2s+1)} \alpha^{2s+1}\right)^{1/2}.
\end{equation}
\par
%
%
{}From equation (\ref{eq:s-g}), it follows that the first inequality in (\ref{eq:condition1}) is
automatically satisfied. However, the second one implies that $l$ may only vary in the range
$l=0$, 1,~\ldots, $l_{\rm max}$, where $l_{\rm max}$ is the largest integer fulfilling the
condition
\begin{equation}
  (l+1)(2l+1) < \frac{2Z}{\alpha} - 1.  \label{eq:condition1bis}
\end{equation}
\par
%
%
Let us next consider the excited state wave functions $R_{kl}(\rho) = R_k(g, s; \rho)$,
$k=1$, 2,~\ldots, which can be determined from the recursion relation
\begin{equation}
  R_{k+1}(g, s; \rho) = [e_{k+1}(g, s) - e_0(g, s)]^{-1/2} B^+(g, s) R_k(g_1, s_1; \rho)
  \label{eq:recursion}
\end{equation}
where $s_1$ and $g_1$ are defined according to equations (\ref{eq:s-i}) and (\ref{eq:g-i}),
respectively. It can be easily shown that the normalizable solutions of equation
(\ref{eq:recursion}) are given by
\begin{equation}
  R_k(g, s; \rho) = {\cal N}_k(g, s) P_k(g, s; \rho) \rho^s (1 +
  \alpha\rho)^{-\left(\frac{g_k}{\alpha} + s_k + \frac{1}{2}\right)}  \label{eq:radial-k}
\end{equation}
where
\begin{equation}
  s_k > 0 \qquad g_k > 0  \label{eq:condition2}
\end{equation}
$P_k(g, s; \rho)$ denotes some $k$th-degree polynomial in $\rho$, satisfying the relation
\begin{equation}
  P_{k+1}(g, s; \rho) = - \rho f(\rho) P'_k(g_1, s_1; \rho) + [- (2s+1) + (g_{k+1} + g
  +k\alpha)\rho] P_k(g_1, s_1; \rho)  \label{eq:polynomial}
\end{equation}
with $P_0(g, s; \rho) \equiv 1$, and ${\cal N}_k(g, s)$ is some normalization coefficient
fulfilling the recursion relation
\begin{eqnarray}
  {\cal N}_{k+1}(g, s) & = & \{2[e_{k+1}(g, s) - e_0(g, s)]\}^{-1/2} {\cal N}_k(g_1, s_1)
       \nonumber \\
  & = & (s+k+1) \left\{(k+1) (2s+k+1) \left[g + \frac{\alpha}{2} (2s+k+1)\right]
       \left[g - \frac{\alpha}{2} (k+1)\right]\right\}^{-1/2}  \nonumber \\
  && \mbox{} \times {\cal N}_k(g_1, s_1).  \label{eq:rel-norm}
\end{eqnarray}
\par
%
%
The second condition in (\ref{eq:condition2}) is equivalent to the inequality
\begin{equation}
  (l+k+1)^2 + l(l+1) < \frac{2Z}{\alpha} - 1  \label{eq:condition2bis}
\end{equation}
generalizing equation (\ref{eq:condition1bis}). It implies that both $l$ and $k$ run over some
finite sets, $l=0$, 1,~\ldots, $l_{\rm max}$ and $k=0$, 1,~\ldots, $k_{\rm max}$. We
therefore conclude that in contrast with the conventional Coulomb problem, the deformed one
corresponding to $f(\rho) = 1 + \alpha\rho$ has only a finite number of bound states.\par
%
%
{}For the first few $k$ values, explicit expressions of the polynomials $P_k(g, s; \rho)$ can be
obtained by solving equation (\ref{eq:polynomial}). For the first two excited states, for
instance, we get
\begin{eqnarray}
  P_1(g, s; \rho) & = & - (s + s_1) + (g + g_1) \rho \\
  P_2(g, s; \rho) & = &  (s + s_1)(s_1 + s_2) - [(s + s_2)(g_1 + g_2) + (s_1 + s_2)(g +
       g_2 + \alpha)] \rho  \nonumber \\
  && \mbox{}  + (g + g_2)(g_1 + g_2) \rho^2.
\end{eqnarray}
\par
%
%
In the $\alpha \to 0$ limit, the radial wave functions obtained in the present section should
give back the conventional ones~\cite{landau}.~\footnote{Contrary to what is done
in~\cite{landau}, we use the conventional definition~\cite{erdelyi} of generalized Laguerre
polynomials in equations (\ref{eq:P-L}), (\ref{eq:rel-L}), and (\ref{eq:radial-wf}).} Recalling
that $s = l+1$ and $g = Z/(l+1)$ in such a limit, we easily get the usual result for the ground
state wave function of $h^{(l)}_1$,
\begin{equation}
  R_{0l}(\rho) = N_{0l} \rho^{l+1} \exp\left(- \frac{Z\rho}{l+1}\right)
\end{equation}
where
\begin{equation}
  N_{0l} = \frac{1}{\sqrt{(2l+2)!}} \left(\frac{2Z}{l+1}\right)^{l+3/2}.  \label{eq:norm-0}
\end{equation}
\par
%
%
{}For the excited state wave functions, it can be shown that for $\alpha \to 0$, the
polynomials $P_k(g, s; \rho)$ become
\begin{equation}
  P_{kl}(\rho) = a_{kl} L^{(2l+1)}_k(t) \qquad t \equiv \frac{2Z\rho}{n} = 
  \frac{2Z\rho}{k+l+1}  \label{eq:P-L}
\end{equation}
where
\begin{equation}
  a_{kl} = (-1)^k \frac{k!\, (2l)!!\, (2k+2l+1)!}{(2k+2l)!!\, (k+2l+1)!}  \label{eq:a-kl}
\end{equation}
and $L^{(2l+1)}_k(t)$ is a generalized Laguerre polynomial. Inserting (\ref{eq:P-L}) in
equation (\ref{eq:polynomial}) where $\alpha$ is set equal to zero, we indeed get the relation
\begin{eqnarray}
  \lefteqn{(2l+2) t \frac{d}{dt} L^{(2l+3)}_{k-1}(t)} \nonumber \\
  && = k (k+2l+2) L^{(2l+1)}_k(t) + [- (2l+2)(2l+3) + (k+2l+2)t] L^{(2l+3)}_{k-1}(t)  
  \label{eq:rel-L}
\end{eqnarray}
which can be proved to hold true from the known relations satisfied by generalized Laguerre
polynomials~\cite{erdelyi}. Finally, on using equations (\ref{eq:rel-norm}), (\ref{eq:norm-0}),
(\ref{eq:P-L}) and (\ref{eq:a-kl}), we obtain that the functions (\ref{eq:radial-k}) lead to
\begin{equation}
  R_{kl}(\rho) = N_{kl} \rho^{l+1} L^{(2l+1)}_k\left(\frac{2Z\rho}{n}\right) \exp\left(-
  \frac{Z\rho}{n}\right)  \label{eq:radial-wf}
\end{equation}
where
\begin{equation}
  N_{kl} = (-1)^{n-l-1} \left(\frac{2Z}{n} \frac{(n-l-1)!}{2n (n+l)!}\right)^{1/2}
  \left(\frac{2Z}{n}\right)^{l+1}.
\end{equation}
\par
%
%
\subsection{Position-dependent mass}

Let us now make the choice (\ref{eq:choice}) with $V_2(r) = V(r)$ in equation (\ref{eq:H-2}). 
The Hamiltonian $H_2$ then describes an electron in a Coulomb potential with a 
position-dependent mass
\begin{equation}
  m(r) = \frac{m_0}{(1 + \aalpha r)^2} \qquad {\rm or} \qquad m(\rho) = \frac{m_0}{(1 +
  \alpha \rho)^2} 
\end{equation}
decreasing from $m_0$ to 0 when the radial variable increases from 0 to $\infty$.\par
%
%
In the associated Schr\"odinger equation (\ref{eq:SE-2}), the effective potential $U_{2,{\rm
eff}}(\rho)$ reads
\begin{eqnarray}
  U_{2,{\rm eff}}(\rho) & = & - \frac{Z^*}{\rho} + \frac{1}{2} \alpha^2 \left[1 - \delta -
  \lambda + \left(\frac{1}{2} - \delta\right) \left(\frac{1}{2} - \lambda\right)\right] \\
  Z^* & \equiv & Z - \frac{\alpha}{2} (1 - \delta - \lambda).  \label{eq:Z*} 
\end{eqnarray}
Apart from some additive constant, it amounts to a Coulomb potential depending on an
effective charge $Z^*$. Hence we can avail ourselves of the results proved in sections 2.2
and 3.1 to provide the solutions $e_{kl}$ (or $e_{nl}$) and $\psi_{klm}(\rho, \theta,
\varphi)$ of equation (\ref{eq:SE-2}).\par
%
%
The spectrum is given by
\begin{eqnarray}
  e_{nl} & = & - \frac{\left[Z - \frac{\alpha}{2}(L^2 + 2 - \delta - \lambda)\right]^2}{2n^2}
       - \frac{\alpha^2}{8} n^2 + \frac{\alpha}{2}\left[Z + \frac{\alpha}{2}(L^2 + \delta +
       \lambda)\right] \nonumber \\
  && \mbox{} + \frac{\alpha^2}{2}\left[1 - \delta - \lambda + \left(\frac{1}{2} - \delta\right)
       \left(\frac{1}{2} - \lambda\right)\right]
\end{eqnarray}
where $n = k+l+1$. From (\ref{eq:condition2bis}) and (\ref{eq:Z*}), it results that the range
of allowed $l$ and $k$ values is now determined by the modified condition
\begin{equation}
  (l+k+1)^2 + l(l+1) < \frac{2Z}{\alpha} - (2 - \delta - \lambda)
\end{equation}
which, apart from $Z$, depends on the mass parameter $\alpha$ and on the parameter
$\kappa = 2 - \delta - \lambda$ related to the mass-ordering ambiguity problem. The
distances between consecutive levels are also entirely governed by these two parameters.\par
%
%
{}Finally, the wave functions $\psi_{klm}(\rho, \theta, \varphi)$ can be obtained from
equations (\ref{eq:psi}) and (\ref{eq:radial-k}) with $Z^*$ substituted for $Z$.\par
%
%
\subsection{Curved space}

Let us finally make the choice (\ref{eq:choice}) with $V_3(r) = V(r)$ in equation
(\ref{eq:H-3}). For the metric tensor $g(\rho) = 1/f^2(\rho) = 1/(1+\alpha\rho)^2$, we
obtain from (\ref{eq:curvature}) that the nonconstant space curvature is given by
\begin{equation}
  R = - 2\alpha \left(\frac{4}{\rho} + \alpha\right)
\end{equation}
and is therefore negative for all $\rho$ values. In such a space, $V_3(r)$, as given in
(\ref{eq:choice}), may not be interpreted as a Coulomb potential since, as shown in the
appendix, the latter assumes there a more complicated form.\par
%
%
In the associated Schr\"odinger equation (\ref{eq:SE-3bis}), the effective potential $U_{3,{\rm
eff}}(\rho)$ reads
\begin{eqnarray}
  U_{3,{\rm eff}}(\rho) & = & - \frac{Z^{**}}{\rho} - \frac{1}{2} \alpha^2 \\
  Z^{**} & \equiv & Z + \alpha.
\end{eqnarray}
Apart from some additive constant, it is therefore similar to $U_3(\rho)$.\par
%
%
The results proved in sections 2.3 and 3.1 lead to the spectrum
\begin{equation}
  e_{nl} = - \frac{\left[Z - \frac{\alpha}{2}(L^2-1)\right]^2}{2n^2} - \frac{\alpha^2}{8} n^2
  + \frac{\alpha}{2}\left[Z + \frac{\alpha}{2}(L^2+1)\right]
\end{equation}
where $n = k+l+1$ and the allowed $k$ and $l$ values are determined by the inequality
\begin{equation}
  (l+k+1)^2 + l(l+1) < \frac{2Z}{\alpha} + 1.
\end{equation}
The corresponding wave functions $\psi_{klm}(\rho, \theta, \varphi)$ are given by equations
(\ref{eq:psi}) and (\ref{eq:radial-k}) with $Z^{**}$ substituted for $Z$.\par
%
%
\section{Conclusion}

In the present paper, we have shown that there exist some intimate connections between
three unconventional Schr\"odinger equations based on the use of some deformed canonical
commutation relations, of a position-dependent effective mass or of a curved space,
respectively. This occurs whenever a specific relation between the deforming function
$f(\xb)$, the position-dependent mass $m(\xb)$ and the (diagonal) metric tensor $g(\xb)$
holds true (see equation (\ref{eq:result})).\par
%
%
As a consequence, any ES Schr\"odinger equation known in one of these three fields can
be reinterpreted as an ES Schr\"odinger equation in the other two. For instance, the
resolution of the three-dimensional harmonic oscillator problem with nonzero minimal
uncertainty in position, carried out in ~\cite{kempf95, cq}, provides us, after interchanging
the role of $x_i$ and $p_i$, with the solution of the three-dimensional harmonic oscillator
problem for a position dependence of the mass given by $m(r) = m_0/(1 + \aalpha r^2)^2$,
where $\aalpha \ge 0$.\par
%
%
Here we have given another illustration of such a type of relations by considering the Coulomb
potential $V(r) = - Ze^2/r$ for a deforming function $f(r) = 1 + \aalpha r$ ($\aalpha \ge 0$)
or a position-dependent mass $m(r) = m_0/(1 + \aalpha r)^2$ or else a similar potential
(then distinct from Coulomb) for a diagonal metric tensor $g(r) = 1/(1 + \aalpha r)^2$. In all
the cases, we have derived the bound-state energy spectrum and the corresponding wave
functions. We have shown that in contrast with the standard case, but in analogy with the
Coulomb potential in a space of constant negative curvature~\cite{schrodinger}, there are
only a finite number of bound states.\par
%
%
It should be stressed that contrary to many constructions of ES Schr\"odinger equations with
position-dependent mass, which start from some known ES problem with constant mass and
then deform the potential while leaving the spectrum unchanged, in our approach we consider
a known ES potential and determine the effect of a mass position dependence on the
spectrum and wave functions.\par
%
%
\section*{Acknowledgments}

CQ is a Research Director of the National Fund for Scientific Research (FNRS), Belgium. VMT
thanks this institution for financial support.\par
%
%
\section*{Appendix. Space curvature and Coulomb potential in curved space}
\renewcommand{\theequation}{A.\arabic{equation}}
\setcounter{section}{0}
\setcounter{equation}{0}

The curvature of the space, whose squared line element is given by equation (\ref{eq:line}),
can be expressed as~\cite{moller}
\begin{equation}
  R = g^{ik} R_{ik}  \label{eq:R}
\end{equation}
in terms of the inverse of the metric tensor $g^{ik}$ and the (contracted) curvature tensor
\begin{equation}
  R_{ik} = \frac{\partial \Gamma_{il}^l}{\partial X^k} - \frac{\partial \Gamma_{ik}^l}{\partial
  X^l} + \Gamma_{il}^r \Gamma_{kr}^l - \Gamma_{ik}^r \Gamma_{lr}^l
\end{equation}
where
\begin{eqnarray}
  \Gamma_{kl}^i & = & g^{im} \Gamma_{m,kl} \\
  \Gamma_{i,kl} & =& \frac{1}{2} \left(\frac{\partial g_{ik}}{\partial X^l}
       + \frac{\partial g_{il}}{\partial X^k} - \frac{\partial g_{kl}}{\partial X^i}\right). 
\end{eqnarray}
\par
%
%
In this appendix, we restrict ourselves to a diagonal metric tensor depending only on $\rho =
(\sum_i X_i^2)^{1/2}$, i.e., $g_{ij} = \delta_{i,j} g(\rho) = \delta_{i,j}/f^2(\rho)$. Then
$g^{ij}$ is given by $g^{ij} = \delta_{i,j}/g(\rho)$ and a straightforward calculation leads to
\begin{eqnarray}
  \Gamma_{kl}^i & = & a(\rho) \left(\delta_{i,k} X^l + \delta_{i,l} X^k - \delta_{k,l} X^i\right)
         \\
  R_{ik} & = & \left(\frac{a'(\rho)}{\rho} - a^2(\rho)\right) X^i X^k + \delta_{i,k} \left[
         4a(\rho) + \rho a'(\rho) + \rho^2 a^2(\rho)\right]  \label{eq:Rik}
\end{eqnarray} 
where the prime denotes derivation with respect to $\rho$ and
\begin{equation}
  a(\rho) = \frac{g'(\rho)}{2\rho g(\rho)}.
\end{equation}
From (\ref{eq:R}) and (\ref{eq:Rik}), we obtain
\begin{equation}
  R = \frac{2}{g(\rho)} \left\{4a(\rho) + 2[\rho a(\rho)]' + \rho^2 a^2(\rho)\right\}
\end{equation}
which can also be rewritten in terms of $f(\rho)$ as shown in equation
(\ref{eq:curvature}).\par
%
%
The Coulomb potential $\phi(\rho)$ in such a curved space can be obtained as a solution of
Laplace equation
\begin{equation}
  \Delta \phi(\rho) = 0  \label{eq:laplace}
\end{equation}
going to $q/\rho$ when $f(\rho) \to 1$ (with $q$ the electric charge).\par
%
%
{}From equation (\ref{eq:laplacian}), it follows that we can rewrite (\ref{eq:laplace}) as
\begin{equation}
  f^2(\rho) \left(\nablab \cdot \nablab - \frac{f'(\rho)}{\rho f(\rho)} \xxb \cdot
  \nablab\right) \phi(\rho) = 0
\end{equation}
or
\begin{equation}
  \left(\frac{d^2}{d\rho^2} + \frac{2}{\rho} \frac{d}{d\rho} - \frac{f'(\rho)}{f(\rho)}
  \frac{d}{d\rho}\right) \phi(\rho) = 0.
\end{equation}
The (nonconstant) solution of this equation reads
\begin{equation}
  \phi(\rho) = C_1 \int^{\rho} d\rho'\, \frac{f(\rho')}{\rho^{\prime2}} + C_2
\end{equation}
where $C_1$ and $C_2$ are two integration constants. For $f(\rho) = 1$, we recover
$\phi(\rho) = q/\rho$ by setting $C_1 = -q$ and $C_2 = 0$. Hence we may choose
\begin{equation}
  \phi(\rho) = -q \int^{\rho} d\rho'\, \frac{f(\rho')}{\rho^{\prime2}}.
\end{equation}
\par
%
%
In the special case $f(\rho) = 1 + \alpha \rho$ considered in section 3.3, the Coulomb
potential is therefore given by
\begin{equation}
  \phi(\rho) = \frac{q}{\rho} - q \alpha \ln\rho.
\end{equation}
%
%


\newpage
\begin{thebibliography}{99}

\bibitem{wannier} Wannier G H 1937 {\sl Phys.\ Rev.} {\bf 52} 191\\
Slater J C 1949 {\sl Phys.\ Rev.} {\bf 76} 1592\\
Luttinger J M and Kohn W 1955 {\sl Phys.\ Rev.} {\bf 97} 869

\bibitem{bastard} Bastard G 1988 {\sl Wave Mechanics Applied to Semiconductor
Heterostructures} (Les Ulis: Editions de Physique)

\bibitem{serra} Serra L and Lipparini E 1997 {\sl Europhys.\ Lett.} {\bf 40} 667

\bibitem{ring} Ring P and Schuck P 1980 {\sl The Nuclear Many Body Problem} (New York:
Springer-Verlag)

\bibitem{arias} Arias de Saavedra F, Boronat J, Polls A and Fabrocini A 1994 {\sl Phys.\ Rev.}
B {\bf 50} 4248

\bibitem{barranco} Barranco M, Pi M, Gatica S M, Hernandez E S and Navarro J 1997 {\sl
Phys.\ Rev.} B {\bf 56} 8997

\bibitem{puente} Puente A, Serra Ll and Casas M 1994 {\sl Z.\ Phys.} D {\bf 31} 283

\bibitem{levy} L\'evy-Leblond J M 1995 {\sl Phys.\ Rev.} A {\bf 52} 1845

\bibitem{cavalcante} Cavalcante F S A, Costa Filho R N, Ribeiro Filho J, de Almeida C A S and
Freire V N 1997 {\sl Phys.\ Rev.} B {\bf 55} 1326

\bibitem{bhatta} Bhattacharjie A and Sudarshan E C G 1962 {\sl Nuovo Cimento} {\bf 25}
864\\
Natanzon G A 1979 {\sl Theor.\ Math.\ Phys.} {\bf 38} 146\\
L\'evai G 1989 {\sl J.\ Phys.\ A: Math.\ Gen.} {\bf 22} 689

\bibitem{alhassid} Alhassid Y, G\" ursey F and Iachello F 1986 {\sl Ann.\ Phys., N.Y.} {\bf
167} 181\\
Wu J and Alhassid Y 1990 {\sl J.\ Math.\ Phys.} {\bf 31} 557\\
Englefield M J and Quesne C 1991 {\sl J.\ Phys.\ A: Math.\ Gen.} {\bf 24} 3557\\
L\'evai G 1994 {\sl J.\ Phys.\ A: Math.\ Gen.} {\bf 27} 3809

\bibitem{gendenshtein} Gendenshtein L E 1983 {\sl JETP Lett.} {\bf 38} 356\\
Cooper F, Khare A and Sukhatme U 1995 {\sl Phys.\ Rep.} {\bf 251} 267\\
Junker G 1996 {\sl Supersymmetric Methods in Quantum and Statistical Physics} (Berlin:
Springer-Verlag)

\bibitem{turbiner} Turbiner A V 1988 {\sl Commun.\ Math.\ Phys.} {\bf 118} 467\\
Shifman M A 1989 {\sl Int.\ J.\ Mod.\ Phys.} A {\bf 4} 3311\\
Ushveridze A G 1994 {\sl Quasi-exactly Solvable Models in Quantum Mechanics} (Bristol: IOP)\\
Bagchi B and Ganguly A 2003 {\sl J.\ Phys.\ A: Math.\ Gen.} {\bf 36} L161

\bibitem{dutra93} de Souza Dutra A 1993 {\sl Phys.\ Rev.} A {\bf 47} R2435\\
Dutt R, Khare A and Varshni Y P 1995 {\sl J.\ Phys.\ A: Math.\ Gen.} {\bf 28} L107\\
Roychoudhury R, Roy P, Znojil M and L\'evai G 2001 {\sl J.\ Math.\ Phys.} {\bf 42} 1996

\bibitem{dekar} Dekar L, Chetouani L and Hammann T F 1998 {\sl J.\ Math.\ Phys.} {\bf 39}
2551\\
Dekar L, Chetouani L and Hammann T F 1999 {\sl Phys.\ Rev.} A {\bf 59} 107

\bibitem{milanovic} Milanovi\'c V and Ikoni\'c Z 1999 {\sl J.\ Phys.\ A: Math.\ Gen.} {\bf 32}
7001

\bibitem{plastino} Plastino A R, Rigo A, Casas M, Garcias F and Plastino A 1999 {\sl Phys.\
Rev.} A {\bf 60} 4318\\
Plastino A R, Puente A, Casas M, Garcias F and Plastino A 2000 {\sl Rev.\ Mex.\ Fis.} {\bf 46}
78

\bibitem{dutra00} de Souza Dutra A and Almeida C A S 2000 {\sl Phys.\ Lett.} A {\bf 275}
25 \\
de Souza Dutra A, Hott M and Almeida C A S 2003 {\sl Europhys.\ Lett.} {\bf 62} 8

\bibitem{roy} Roy B and Roy P 2002 {\sl J.\ Phys.\ A: Math.\ Gen.} {\bf 35} 3961

\bibitem{koc} Ko\c c R, Koca M and K\"orc\"uk E 2002 {\sl J.\ Phys.\ A: Math.\ Gen.} {\bf 35}
L527\\
Ko\c c R and Koca M 2003 {\sl J.\ Phys.\ A: Math.\ Gen.} {\bf 36} 8105

\bibitem{alhaidari} Alhaidari A D 2002 {\sl Phys.\ Rev.} A {\bf 66} 042116

\bibitem{gonul} G\"on\"ul B, G\"on\"ul B, Tutcu D and \"Ozer O 2002 {\sl Mod.\ Phys.\
Lett.} A {\bf 17} 2057\\
G\"on\"ul B, \"Ozer O, G\"on\"ul B and \"Uzg\"un F  2002 {\sl Mod.\ Phys.\ Lett.} A {\bf 17}
2453

\bibitem{kempf94} Kempf A 1994 {\sl J.\ Math.\ Phys.} {\bf 35} 4483 \\
Hinrichsen H and Kempf A 1996 {\sl J.\ Math.\ Phys.} {\bf 37} 2121\\
Kempf A 1997 {\sl J.\ Phys.\ A: Math.\ Gen.} {\bf 30} 2093

\bibitem{witten} Witten E 1996 {\sl Phys.\ Today} {\bf 49} 24

\bibitem{kempf95} Kempf A, Mangano G and Mann R B 1995 {\sl Phys.\ Rev.} D {\bf 52}
1108\\ 
Chang L N, Minic D, Okamura N and Takeuchi T 2002 {\sl Phys.\ Rev.} D {\bf 65} 125027

\bibitem{cq} Quesne C and Tkachuk V M 2003 {\sl J.\ Phys.\ A: Math.\ Gen.} {\bf 36}
10373\\
Quesne C and Tkachuk V M 2003 More on a SUSYQM approach to the harmonic oscillator with
nonzero minimal uncertainties in position and/or momentum {\sl Preprint} math-ph/0312029

\bibitem{schrodinger} Schr\"odinger E 1941 {\sl Proc.\ R.\ Irish Acad.} {\bf 46} 183\\
Stevenson A F C 1941 {\sl Phys.\ Rev.} {\bf 59} 842\\
Infeld L and Schild A 1945 {\sl Phys.\ Rev.} {\bf 67} 121

\bibitem{higgs} Higgs P W 1979 {\sl J.\ Phys.\ A: Math.\ Gen.} {\bf 12} 309\\
Leemon H I 1979 {\sl J.\ Phys.\ A: Math.\ Gen.} {\bf 12} 489

\bibitem{granovskii} Granovskii Ya I, Zhedanov A S and Lutsenko I M 1992 {\sl Theor.\ Math.\
Phys.} {\bf 91} 474\\
Granovskii Ya I, Zhedanov A S and Lutsenko I M 1992 {\sl Theor.\ Math.\ Phys.} {\bf 91} 604

\bibitem{kalnins96} Kalnins E G, Miller Jr W and Pogosyan G S 1996 {\sl J.\ Math.\ Phys.} {\bf
37} 6439\\
Kalnins E G, Miller Jr W and Pogosyan G S 1997 {\sl J.\ Math.\ Phys.} {\bf 38} 5416

\bibitem{kalnins02} Kalnins E G, Kress J M and Winternitz P 2002 {\sl J.\ Math.\ Phys.} {\bf
43} 970\\
Kalnins E G, Kress J M, Miller Jr W and Winternitz P 2003 {\sl J.\ Math.\ Phys.} {\bf 44} 5811

\bibitem{vonroos} von Roos O 1983 {\sl Phys.\ Rev.} B {\bf 27} 7547

\bibitem{bendaniel} BenDaniel D J and Duke C B 1966 {\sl Phys.\ Rev.} B {\bf 152} 683

\bibitem{gradshteyn} Gradshteyn I S and Ryzhik I M 1980 {\sl Table of Integrals, Series, and
Products} (New York: Academic)

\bibitem{landau} Landau L D and Lifshitz E M 1958 {\sl Quantum Mechanics} (Oxford:
Pergamon)

\bibitem{erdelyi} Erd\'elyi A, Magnus W, Oberhettinger F and Tricomi F G 1953 {\sl Higher
Transcendental Functions} vol II (New York: McGraw-Hill)

\bibitem{moller} M\o ller C 1952 {\sl The Theory of Relativity} (Oxford: Clarendon)

\end {thebibliography}
\end{document}